\documentclass[%
 reprint,
 amsmath,amssymb,
 aps,
]{revtex4-1}
\usepackage{bbm}
\usepackage{graphicx}% Include figure files
\usepackage{dcolumn}% Align table columns on decimal point
\usepackage{bm}% bold math
\makeatletter
\newcommand{\rmnum}[1]{\romannumeral #1}
\newcommand{\Rmnum}[1]{\expandafter\@slowromancap\romannumeral #1@}
\makeatother

\begin{document}

\preprint{APS/123-QED}

\title{How to enhance the dynamic range of excitatory-inhibitory excitable networks}% Force line breaks with \\

\author{Sen Pei}
\email{peisen@smss.buaa.edu.cn}
\author{Shaoting Tang}
\author{Shu Yan}
\author{Shijin Jiang}
\altaffiliation[Also at ]{School of Mathematical Sciences, Peking
University.}
\author{Xiao Zhang}
\author{Zhiming Zheng}%
 \email{zzheng@pku.edu.cn}
\affiliation{%
Key Laboratory of Mathematics, Informatics and Behavioral Semantics,
Ministry of Education\\School of Mathematics and Systems Science,
Beihang University
}%

\date{\today}% It is always \today, today,

\begin{abstract}
We investigate the collective dynamics of excitatory-inhibitory
excitable networks in response to external stimuli. How to enhance
dynamic range, which represents the ability of networks to encode
external stimuli, is crucial to many applications. We regard the
system as a two-layer network (E-Layer and I-Layer) and explore the
criticality and dynamic range on diverse networks. Interestingly, we
find that phase transition occurs when the dominant eigenvalue of
E-layer's weighted adjacency matrix is exactly one, which is only
determined by the topology of E-Layer. Meanwhile, it is shown that
dynamic range is maximized at critical state. Based on theoretical
analysis, we propose an inhibitory factor for each excitatory node.
We suggest that if nodes with high inhibitory factors are cut out
from I-Layer, dynamic range could be further enhanced. However,
because of the sparseness of networks and passive function of
inhibitory nodes, the improvement is relatively small compared to
original dynamic range. Even so, this provides a strategy to enhance
dynamic range.

\end{abstract}

\pacs{Valid PACS appear here}% PACS, the Physics and Astronomy
\maketitle

%\tableofcontents

\section{\label{sec1}Introduction}

In last decades, the theory of complex networks \cite{WS,AB} has
enjoyed tremendous development in many fields as diverse as neural
science \cite{KC,BS}, epidemic control \cite{PV}, social activities
\cite{CFL}, economics \cite{YR,PTZ}, etc. Especially, in the
research of neural networks, many enlightening theoretical results,
which are verified by experimental systems, are obtained. For
example, Beggs \emph{et al} \cite{BP,BP1} shows that neural
avalanches, which have a power law size distribution, are important
for cortical information processing and storage; Soriano \emph{et
al} \cite{SMTM,BSMT} relates neural cultures with percolation on a
graph, obtaining a percolation transition in connectivity
characterized by a power law.

In applications, many biological \cite{KC}, social \cite{ZK} and
engineering problems \cite{FTV} are accurately described as networks
of coupled excitable systems. The studies of how such networks
respond to external stimuli reveal that, although single nodes
usually respond to stimuli with small ranges, the collective
response of the entire system can encode stimuli spanning several
orders of magnitude. This property of broad dynamic range (the range
of stimulus intensities resulting in distinguishable network
response) is of particular significance for information processing
in sensory neural networks \cite{CROK,CORK}.

In order to explain this phenomenon, a model of an excitable network
based on \(Erd\ddot{o}s-R\acute{e}nyi\) random graphs is proposed by
Kinouchi and Copelli \cite{KC}. It is shown that such networks have
their sensitivity and dynamic range maximized at the critical point
of a non-equilibrium phase transition \cite{KC}. Later on, models on
other diverse networks including those with scale-free,
degree-correlated, and assortative topologies are discussed
 \cite{CC,WXW}. More recently, a general theoretical approach to
study the effects of network topology on dynamic range is presented
by Larremore, Shew and Restrepo \cite{LSR,LSOR}. Interestingly,
results show that the dynamic range is governed by the largest
eigenvalue of the weighted network adjacency matrix.

All the models discussed above have only considered excitatory
nodes. However, in realistic neural systems, the excitatory and
inhibitory neurons are coexisting \cite{AST}. Such
excitatory-inhibitory (E-I) networks arise in many regions
throughout the central nervous system and display complex patterns
of activity \cite{PT,FE}. The behavior of E-I networks is critical
for understanding how neural circuits produce cognitive function
 \cite{VRA}. Therefore, huge effort on E-I networks has been made. It
is shown that excitation/inhibition balance is crucial for
transmission of rate code in long feedforward networks
\cite{LSSA,SN0}, signal propagation in spiking neural networks
\cite{KAK,VA} and discharge of cortical neurons \cite{SN}. Although
some other profound studies on E-I networks are presented
\cite{B,OL,WC}, how such networks respond to external stimuli is
largely unknown. Moreover, the investigation for the criticality and
dynamic range of E-I networks with arbitrary topology still remains
open.

In this paper, we investigate the criticality and dynamic range of
E-I network models, providing a strategy to further enhance dynamic
range. In section \ref{sec2}, we propose the excitatory-inhibitory
network model and give some basic definitions about criticality and
dynamic range. Then in section \ref{sec3} and \ref{sec4}, we conduct
a theoretical study on ER random networks and scale-free networks
respectively. It is proved that the criticality only relates to the
topology of E-layer. In section \ref{sec5}, we investigate methods
to further enhance dynamic range of given networks by analyzing the
mutual effects of E-layer and I-layer. We analyze the upper bound on
improvement and explain why it is small compared to original dynamic
range. Lastly, in section \ref{sec6}, we conclude the results and
give a discussion about the further research.

\section{\label{sec2}The Excitatory-Inhibitory Network Model}

\begin{figure}
  \includegraphics[width=3.5in]{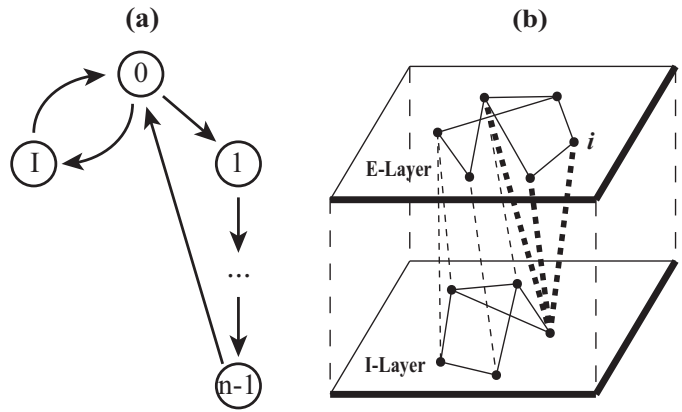}\\
  \caption{Illustrations of the state of excitable nodes and E-I layered network.
(a) The state evolution of an excitable node. The symbol I
represents the inhibited state. (b) A schema of a E-I network. The
E-layer is consist of excitatory nodes while the I-layer only
contains inhibitory nodes. The bold dotted lines are inhibitory
paths of node \(i\).}\label{illustration}
\end{figure}

In the present model, each excitable element \(i=1,\ldots,N\) has
\(n\) states: \(s_{i}=0\) is the resting state, \(s_{i}=1\)
corresponds to excitation and the remaining \(s_{i}=2,\ldots,n-1\)
are refractory states. Here two types of nodes are considered:
excitatory and inhibitory nodes. The function of excitatory nodes is
to transmit excitation signals, increasing the probability of
excitation of their neighbors, while the inhibitory ones decrease
this probability. To be precise, at discrete times \(t=0,1,\ldots\)
the states of the nodes \(s_{i}^{t}\) are updated as follows:
(\rmnum{1}) If node \(i\) is in the resting state, \(s_{i}^{t}=0\),
it can be inhibited by another excited inhibitory neighbor \(j\),
\(s_{j}^{t}=1\) with probability \(A_{ij}\). In this case, the state
of node \(i\) will remain \(0\) in the next time step. Otherwise, it
can be excited by its excited excitatory neighbor \(j'\) with
probability \(A_{ij'}\), or independently by an external stimulus
with probability \(\eta\). (\rmnum{2}) The dynamics of the nodes
that are excited or in a refractory state is deterministic: if
\(s_{i}=1\), then in the next time step its state changes to
\(s_{i}=2\), and so on until the state \(s_{i}=n-1\) leads to the
\(s_{i}=0\) resting state, see Fig.\(\ref{illustration}\)(a).

The network topology and strength of interactions between the nodes
are described by the weighted adjacency matrix \(A=\{A_{ij}\}\).
Notice that this matrix contains the information of both the
excitatory and inhibitory links. In order to simplify the analysis,
we only consider undirected networks here, so \(A\) is symmetrical.
Besides, in this model, \(\eta\) will be assumed to be proportional
to the stimulus level. Each element receives external signals
independently.

Considering the two distinct types of nodes in this model, we can
regard the system as a layered network \cite{KT}. The upper layer is
composed of excitatory nodes, while the lower layer only contains
inhibitory ones. We denote them as E-layer and I-layer respectively.
An illustration of this two-layer model is shown in
Fig.\(\ref{illustration}\)(b). Assume the E-layer has \(N_{e}\)
nodes, and I-layer has \(N_{i}\) nodes, then \(N_{e}+N_{i}=N\).
Denote \(f_{e}\) and \(f_{i}\) as the fraction of excitatory and
inhibitory nodes. We have \(f_{e}=N_{e}/N\), \(f_{i}=N_{i}/N\). For
the convenience of analysis, we rearrange the indices of nodes so
that elements with index \(1\leq i\leq N_{e}\) represent excitatory
nodes and the others are inhibitory ones. Therefore, we obtain the
following weighted adjacency matrix
\begin{equation}
A_{N\times N}= \left(
  \begin{array}{cc}
    A^{EE} & A^{IE} \\
    A^{EI} & A^{II} \\
  \end{array}
\right).\label{E1}
\end{equation}
Here \(A^{EE}=\{A^{EE}_{ij}\}_{N_{e}\times N_{e}}\) describes the
topology and interaction strength of E-layer.
\(A^{IE}=\{A^{IE}_{ij}\}_{N_{e}\times N_{i}}\) represents the effect
of I-layer on E-layer. The other two matrices have similar meanings.

To analyze the dynamics of this system, we denote the probability
that a given node \(i\) is excited \((s=1)\) at time \(t\) by
\(p_{i}^{t}\). In this model, we define the network instantaneous
activity \(p^{t}\) at time \(t\) as the fraction of excited
excitatory nodes, i.e.
\(p^{t}=\frac{1}{N_{e}}\sum_{i=1}^{N_{e}}p_{i}^{t}\). Notice that we
only care about the excitatory nodes here. We also define the
average activity \(F=\frac{1}{T}\sum_{t=1}^{T}p^{t}\), where \(T\)
is a large time window. As a function of the stimulus intensity
\(\eta\), networks have a minimum response \(F_{0}\) and a maximum
response \(F_{max}\). We define the dynamic range
\(\Delta=10\log(\eta_{high}/\eta_{low})\) as the range of stimuli
that is distinguishable based on the system's response \(F\),
discarding stimuli that are too weak to be distinguished from
\(F_{0}\) or too close to saturation. The range \([\eta_{low},
\eta_{high}]\) is found from its corresponding response interval
\([F_{low}, F_{high}]\), where \(F_{x}=F_{0}+x(F_{max}-F_{0})\). The
choice of interval is arbitrary and does not affect our results.

In the following text, we will investigate this model on networks
with various topologies. As typical examples of homogeneous and
heterogeneous networks, random networks and scale-free networks will
be analyzed separately.

\section{\label{sec3}Random Networks}

The network with \(N\) nodes is an \(Erd\ddot{o}s-R\acute{e}nyi\)
(ER) undirected random graph, with \(NK/2\) links being assigned to
randomly chosen pairs of nodes. This produces an average degree
\(K\). We randomly choose \(N_{e}=f_{e}N\) nodes as excitatory
elements, and the rest as inhibitory ones. For simplicity, we set
the strength of interactions between the nodes as \(S\) uniformly.
In a mean-field approximation, the average branching ratio \cite{KC}
of excitatory nodes \(\sigma=KS\) corresponds to the average number
of excitations created in the next time step by an excited
excitatory element. Of all these excited nodes, \(f_{e}\sigma\)
nodes are excitatory and the others are inhibitory.

In the case of random networks, the update equations for both
E-layer and I-layer are same. Assume that the events of neighbors of
a node being excited at time \(t\) are statistically independent. It
is shown that this approximation yields good results even the
network has a non-negligible amount of short loops \cite{LSR}.
Therefore, we obtain the following mean-field map for \(p^{t}\) at
sufficient long time \(t\):
\begin{eqnarray}
p^{t+1}=(1-(n-1)p^{t})(1-Sp^{t})^{f_{i}K}\nonumber\\
\times \{\eta+(1-\eta)[1-(1-Sp^{t})^{f_{e}K}]\}.\label{E2}
\end{eqnarray}

In the stationary state, \(p^{t+1}=p^{t}=p\). Thus
\(F=\frac{1}{T}\sum_{t=1}^{T}p^{t}\approx p\) for large \(T\). To
check the critical behavior without an external field, we set
\(\eta=0\) and linearize the term \((1-Sp^{t})^{f_{i}K}\) and
\((1-Sp^{t})^{f_{e}K}\) in Eq.(\ref{E2}) around \(p^{t}=0\),
obtaining
\begin{equation}
p\approx (1-(n-1)p)(1-f_{i}\sigma p)f_{e}\sigma p.\label{E3}
\end{equation}
Up to first order, we have the nonzero solution
\begin{equation}
p\approx \frac{\sigma-1/f_{e}}{\sigma(f_{i}\sigma+n-1)}.\label{E4}
\end{equation}
Therefore, the critical point is \(\sigma_{c}=\frac{1}{f_{e}}\),
corresponding to the condition that the average number of
excitations in E-layer caused by an excited node at each time step
is exactly one. In particular, \(\lim_{\eta\rightarrow0}F=0\) if
\(\sigma<\sigma_{c}\) and \(\lim_{\eta\rightarrow0}F>0\) if
\(\sigma>\sigma_{c}\).

Next we will analyze the effect of a vanishing field at the critical
point. In the limit of \(p\rightarrow0\), Eq.(\ref{E2}) can be
approximated by
\begin{equation}
p=(1-(n-1)p)e^{-f_{i}\sigma p}\{\eta+(1-\eta)(1-e^{-f_{e}\sigma
p})\}.\label{E5}
\end{equation}
At the critical point \(\sigma_c=1/f_{e}\), we expand Eq.(\ref{E5})
to second order in the case of \(\eta\rightarrow0\). Then we have
\begin{equation}
p\approx\sqrt{\frac{\eta}{\sigma+n-3/2}}.\label{E6}
\end{equation}

Making use of Eq.(\ref{E5}), we can also predict the dynamic range.
We can solve stimulus \(\eta\) for any given response \(F\). For a
system with refractory time \(n\), the maximal response
\(F_{max}=1/n\). Once we find \(F_{0}\), \(F_{low}\) can be
obtained. Then we can get \(\eta_{low}\) with Eq.(\ref{E5}). For
\(\eta_{high}\), Eq.(\ref{E5}) is invalid because it is only valid
for \(p\rightarrow0\). In order to approximate \(\eta_{high}\), we
set \(\eta_{high}\) to one. Therefore, we have
\begin{equation}
\Delta=-10\log[1-e^{f_{e}\sigma
F_{low}}+\frac{F_{low}}{1-(n-1)F_{low}}e^{\sigma
F_{low}}].\label{E7}
\end{equation}

\begin{figure}
  \includegraphics[width=3.5in]{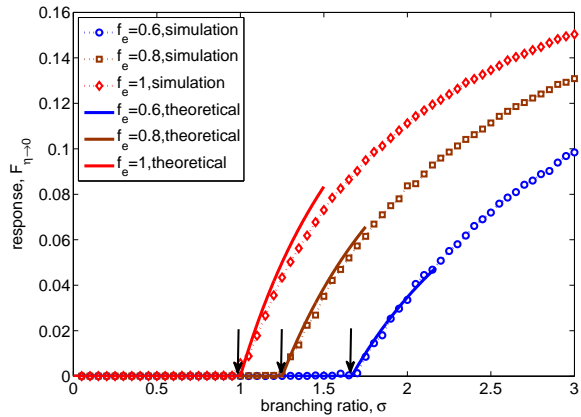}\\
  \caption{Response \(F_{\eta\rightarrow0}\) versus branching ratio \(\sigma\) on ER random networks with \(N=10^4\), \(K=10\), and \(n=5\).
 The solid lines are theoretical values and the symbols represent simulation results. Simulations with \(f_e=0.6\), \(0.8\) and \(1\) are
 plotted with different symbols respectively. The arrows point to critical points of theoretical results.}\label{F_vs_sigma}
\end{figure}

Now we check the critical point of phase transition via simulations.
Fig.\(\ref{F_vs_sigma}\) shows the relationship between response
\(F_{\eta\rightarrow0}\) and branching ratio \(\sigma\) for
different excitatory proportions \(f_e=0.6\), \(0.8\) and \(1\). The
critical point of each situation is in accordance with the
theoretical value, satisfying \(\sigma_c=1/f_e\). Meanwhile, the
behavior near the critical point is well captured by
Eq.\((\ref{E4})\).

\begin{figure}
  \includegraphics[width=3.5in]{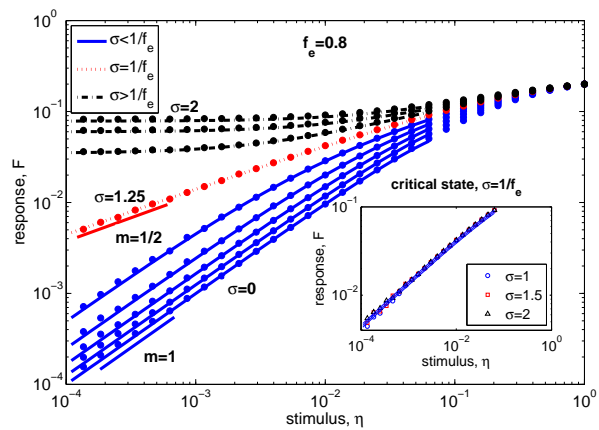}\\
  \caption{Response \(F\) versus stimulus \(\eta\) from \(\sigma=0\) to \(2\) in intervals of \(0.25\) for \(f_e=0.8\).
The points represent simulation results on ER random networks with
\(N=10^4\), \(K=10\), and \(n=5\). The lines correspond to
theoretical result from Eq.\((\ref{E5})\). The line segments show
the power law exponent \(m\). The inset presents the critical state
of \(\sigma=1\), \(1.5\) and \(2\). The lines are the results of
Eq.\((\ref{E6})\). }\label{response_stimuli}
\end{figure}

Then we check the effect of a vanishing field on the system. The
response curves from \(\sigma=0\) to \(2\) (in intervals of
\(0.25\)) for \(f_e=0.8\) are presented in
Fig.\(\ref{response_stimuli}\). The theoretical prediction of
Eq.\((\ref{E5})\) fits the simulation data well. The phase
transition occurs at \(\sigma=1.25\). In critical regime, the power
law exponent \(m=1/2\), compatible with Eq.\((\ref{E6})\).
Meanwhile, in subcritical regime, \(m=1\). The inset of
Fig.\(\ref{response_stimuli}\) shows the critical state for
different branching ratios. The theoretical lines from
Eq.\((\ref{E6})\) agree with the simulations.

\begin{figure}
  \includegraphics[width=3.5in]{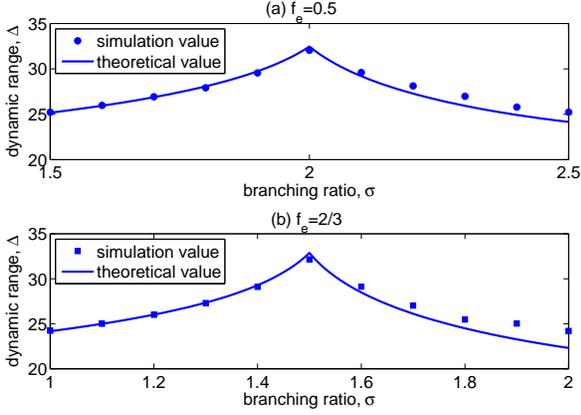}\\
  \caption{Dynamic range \(\Delta\) versus branching ratio \(\sigma\). The points represent simulation results on ER random networks with
\(N=10^4\), \(K=10\), and \(n=5\). The lines correspond to
theoretical results from Eq.\((\ref{E7})\). (a), \(f_e=0.5\),
\(\sigma_c=2\). (b), \(f_e=2/3\), \(\sigma_c=1.5\).
}\label{DR_vs_sigma}
\end{figure}

Fig.\(\ref{DR_vs_sigma}\) shows the dynamic range for \(f_e=0.5\)
and \(2/3\). As can be seen, dynamic range versus branching ratio is
optimized at the critical point \(\sigma_c=1/f_e\). In the
subcritical region, sensitivity is enlarged because weak stimuli are
amplified due to activity propagation among neighbors. Therefore,
the dynamic range increases monotonically with \(\sigma\). In the
supercritical region, the response \(F\), which is positive, masks
the present of weak stimuli, decreasing dynamic range.

Although the mean-field analysis applies for ER random networks
quiet well, it is powerless when tackle the study of heterogeneous
networks, which are ubiquitous in practical complex systems. Thus in
next section, we will mainly deal with the scale-free networks,
especially, the \(Barab\acute{a}si-Albert\) (BA) networks \cite{AB}.

\section{\label{sec4}Scale-Free Networks}

For networks with heterogeneous topology, we have to analyze each
node separately. Assume the connectivity matrix is
\(A=\{A_{ij}\}_{N\times N}\). Then the update equation for
\(p_{i}^{t}\) is
\begin{eqnarray}
p_{i}^{t+1}=(1-(n-1)p_{i}^{t})\prod_{j=N_{e}+1}^{N}(1-A_{ij}p_{j}^{t})\nonumber\\
\times
\{\eta+(1-\eta)[1-\prod_{j=1}^{N_{e}}(1-A_{ij}p_{j}^{t})]\},\label{E8}
\end{eqnarray}

To examine the critical point, we set \(\eta=0\) and expand
Eq.\((\ref{E8})\) to first order in the limit of
\(p_{i}^{t}\rightarrow0\). We have
\begin{equation}
p_{i}^{t+1}=\sum_{j=1}^{N_{e}}A_{ij}p_{j}^{t}.\label{E9}
\end{equation}
Based on matrix \(A\), we create new matrices \(A^{E}\) and
\(A^{I}\) as follows
\begin{equation}
A^{E}= \left(
  \begin{array}{cc}
    A^{EE} & \bm{0} \\
    A^{EI} & \bm{0} \\
  \end{array}
\right),\label{E10}
\end{equation}
\begin{equation}
A^{I}= \left(
  \begin{array}{cc}
    \bm{0} & A^{IE} \\
    \bm{0} & A^{II} \\
  \end{array}
\right).\label{E11}
\end{equation}
Thus, Eq.\((\ref{E9})\) can be transformed into
\begin{equation}
\bm{p}^{t+1}=A^{E}\bm{p}^{t},\label{E12}
\end{equation}
where \(\bm{p}^{t}=(p_{1}^{t},\cdots,p_{N}^{t})^{T}\) is the
probability vector at time \(t\). We can see the stability of the
solution \(\bm{p}=0\) is governed by the dominant eigenvalue
\(\lambda\) of matrix \(A^{E}\). Therefore, the phase transition
happens at \(\lambda_c=1\). Notice that \(A^{E}_{ij}\geq0\), so
\(\lambda\) is real and positive according to Perron-Frobenius
theorem \cite{M}. Recall Eq.\((\ref{E10})\), \(\lambda\) is in fact
the dominant eigenvalue of matrix \(A^{EE}\). This means that the
critical point only determined by the topology of E-layer.

Relating Eq.\((\ref{E12})\) with the power method in numerical
analysis, we find that for small \(\bm{p}\) and \(\eta\), \(\bm{p}\)
should be almost proportional to the normalized right eigenvector
\(\bm{u}\) of \(A^{E}\) corresponding to \(\lambda\). Thus we assume
\(p_{i}=Cu_{i}+\epsilon_{i}\) where \(C\) is a constant and
\(\epsilon_{i}\) is an error term. Based on this, we obtain
\begin{equation}
F=\frac{1}{N_{e}}\sum_{i=1}^{N_{e}}p_{i}\approx\frac{1}{N_{e}}\sum_{i=0}^{N_{e}}Cu_{i}=C\langle
u\rangle_{e}.\label{E13}
\end{equation}
Here \(\langle u\rangle_{e}=\sum_{i=1}^{N_{e}}u_{i}/N_{e}\). Near
the critical state, we can approximate the product terms of
Eq.\((\ref{E8})\) with exponential ones. Then in the steady state,
we have
\begin{eqnarray}
p_{i}=(1-(n-1)p_{i})\exp(-\sum_{j=N_{e}+1}^{N}A_{ij}p_{j})\nonumber\\
\times
\{\eta+(1-\eta)[1-\exp(-\sum_{j=1}^{Ne}A_{ij}p_{j})]\}.\label{E14}
\end{eqnarray}
To solve the stationary state, set \(\eta=0\). Then using
\(p_{i}=Cu_{i}+\epsilon_{i}\) and \(A^{E}\bm{u}=\lambda \bm{u}\), we
expand Eq.\((\ref{E14})\) to second order for \(p_{i}\rightarrow0\).
\begin{eqnarray}
Cu_{i}+\epsilon_{i}=(A^{E}\bm{\epsilon})_{i}+C\lambda
u_{i}-C^{2}\lambda u_{i}(A^{I}\bm{u})_{i}\nonumber\\
-((n-1)\lambda+\frac{1}{2}\lambda^{2})C^{2}u_{i}^{2}.\label{E15}
\end{eqnarray}
In order to eliminate the error term \(\epsilon_{i}\), we make use
of the left eigenvector \(\bm{v}\) corresponding to \(\lambda\)
\cite{LSR}. Recall Eq.\((\ref{E10})\) and \((\ref{E11})\), we can
reveal more details about \(\bm{u}\) and \(\bm{v}\) and simplify the
calculation. Assume the right and left eigenvector of \(A^{EE}\)
corresponding to the dominant eigenvalue \(\lambda\) is
\(\bm{u}^{E}\) and \(\bm{v}^{E}\) respectively. We can express
\(\bm{u}\) and \(\bm{v}\) using \(\bm{u}^{E}\) and \(\bm{v}^{E}\):
\begin{equation}
\bm{u}=(\bm{u}^{E},\frac{1}{\lambda}A^{EI}\bm{u}^{E})^{T},
\bm{v}=(\bm{v}^{E},\bm{0}).\label{E16}
\end{equation}
We multiply Eq.\((\ref{E15})\) by \(v_{i}\) and sum over \(i\).
Using that \(\bm{v}A^{E}\bm{\epsilon}=\lambda\bm{v}\bm{\epsilon}\),
we can neglect the term \((1-\lambda)\sum_{i}v_{i}\epsilon_{i}\) for
\(\lambda\) close to 1. Notice that \(v_{i}=0\) for \(N_{e}+1\leq
i\leq N\). So in fact only the first \(N_{e}\) equations exists.
Thus, we divide the summation by \(N_{e}\). With this equation, the
constant \(C\) can be easily solved. Therefore for \(\eta=0\) the
nonzero solution for \(F\) is
\begin{equation}
F_{\eta=0}=\frac{(\lambda-1)\langle uv\rangle_{e}\langle
u\rangle_{e}}{\lambda((n-1)+\frac{1}{2}\lambda)\langle
u^{2}v\rangle_{e}+\langle u(A^{I}u)v\rangle_{e}},\label{E17}
\end{equation}
where \(\langle uv\rangle_{e}=\sum_{i=1}^{N_{e}}u_{i}v_{i}/N_{e}\),
\(\langle
u^{2}v\rangle_{e}=\sum_{i=1}^{N_{e}}u_{i}^{2}v_{i}/N_{e}\), and
\(\langle
u(A^{I}u)v\rangle_{e}=\frac{1}{N_{e}}\sum_{i=1}^{N_{e}}v_{i}u_{i}\sum_{k}\sum_{j}A_{ij}^{IE}A_{jk}^{EI}u_{k}\).

\begin{figure}
  \includegraphics[width=3.5in]{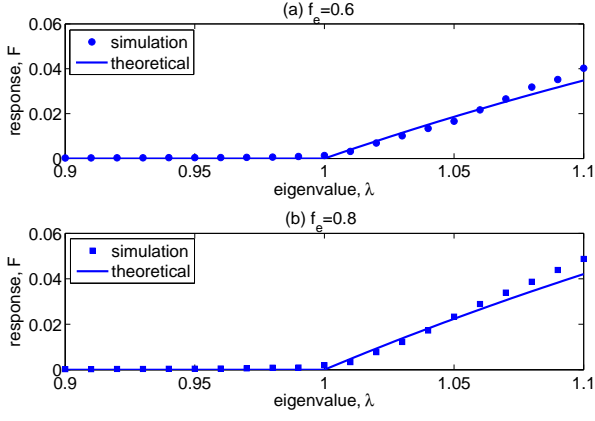}\\
  \caption{Response \(F\) versus the dominant eigenvalue \(\lambda\) of E-layer on BA scale-free networks with node number \(N=10^4\),
mean degree \(\langle k\rangle=4\), and refractory time \(n=2\). The
points represent simulation results, whereas the lines correspond to
the theoretical value from Eq.\((\ref{E17})\). (a), the fraction of
excitatory nodes \(f_e=0.6\).(b),the fraction of excitatory nodes
\(f_e=0.8\). }\label{F_eig}
\end{figure}

In order to test these theoretical results via simulations, we first
created binary networks (\(A_{ij}\in\{0,1\}\)) with the
\(Barab\acute{a}si-Albert\) model. Then we calculate the largest
eigenvalue \(\lambda\) of the binary network and multiply \(A\) by a
constant to rescale the largest eigenvalue to the targeted one.

Fig.\(\ref{F_eig}\) presents the relationship between response
\(F_{\eta\rightarrow0}\) and the largest eigenvalue \(\lambda\) for
\(f_e=0.6\) and \(f_e=0.8\). For both cases, the criticality occurs
at \(\lambda_c=1\). And the behavior near critical point can be
approximated by Eq.\((\ref{E17})\).

\begin{figure}
  \includegraphics[width=3.5in]{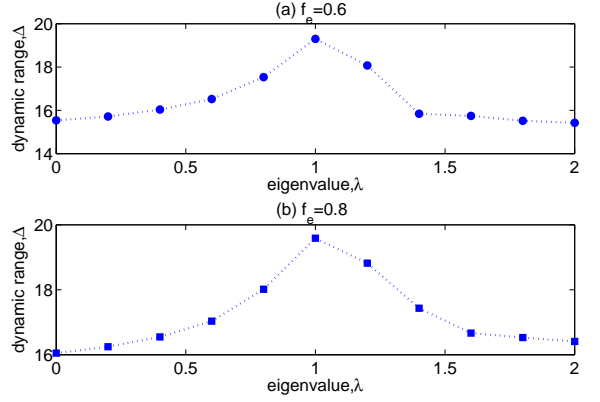}\\
  \caption{Dynamic range \(\Delta\) versus the dominant eigenvalue \(\lambda\) of E-layer on BA scale-free networks with node number \(N=10^4\),
mean degree \(\langle k\rangle=4\), and refractory time \(n=2\). The
points represent simulation results. (a), the fraction of excitatory
nodes \(f_e=0.6\).(b),the proportion of fraction nodes \(f_e=0.8\).
}\label{DR_eig}
\end{figure}

The maximal dynamic range is observed at criticality
\(\lambda_c=1\), see Fig.\(\ref{DR_eig}\). Just as the case of ER
random networks, in the subcritical regime, dynamic range increases
monotonically with the dominant eigenvalue \(\lambda\) due to
enhancement of link strength. On the contrary, in supercritical
region, dynamic range decreases because of the increase of
self-sustained activity \(F\).

Relating to the result of ER random networks, assume the mean degree
of E-layer is \(\langle k\rangle_e\). Because the E-layer and
I-layer are randomly connected, and the fraction of excitatory nodes
is \(f_e\), we have \(\langle k\rangle_e=Kf_e\). In the case of ER
random networks, the largest eigenvalue of \(A^{EE}\) can be
approximated by \(\lambda\approx\langle d\rangle_e\), where
\(\langle
d\rangle_e=\frac{1}{N_e}\sum_{i,j=1}^{N_e}A_{ij}^{EE}\approx
S\langle k\rangle_e=\sigma f_e\). Then the critical point
\(\sigma_c\) satisfies \(\sigma_c f_e=1\), which is identical to the
result of Eq.\((\ref{E4})\).

In the next section, we will explore the way to further enhance
dynamic range of a given network.

\section{\label{sec5}Enhancing Dynamic Range of E-I Networks}

In this section, we will explore the method to enhance dynamic range
of E-I networks. Based on our analysis above, dynamic range is
maximized at criticality. Therefore, it is trivial to adjust the
link strength to make the system reach critical state. Then the
question we face now is how to further improve dynamic range for a
given network at criticality.

Now we give an analysis about the dynamic range at criticality.
Since the dynamic range is maximized at criticality, we set
\(\lambda=1\) in Eq.\((\ref{E14})\). Then we can solve the stimulus
level \(\eta\) corresponding to the response \(F\). To the first
order, we obtain the rough result
\begin{equation}
\eta\approx\frac{(n-\frac{1}{2})\langle u^{2}v\rangle_{e}+\langle
u(A^{I}u)v\rangle_{e}}{\langle v\rangle_{e}\langle
u\rangle_{e}^{2}}F^{2}.\label{E18}
\end{equation}
Here \(\langle
u^{2}v\rangle_{e}=\sum_{i=1}^{N_{e}}u_{i}^{2}v_{i}/N_{e}\) and
\(\langle
u(A^{I}u)v\rangle_{e}=\frac{1}{N_{e}}\sum_{i=1}^{N_{e}}v_{i}u_{i}\sum_{k}\sum_{j}A_{ij}^{IE}A_{jk}^{EI}u_{k}\).
The eigenvector \(\bm{u}\) and \(\bm{v}\) only relate to \(A^{EE}\).
Once \(A^{EE}\) reaches critical state, \(\bm{u}\) and \(\bm{v}\)
will be fixed. However, we can adjust the matrix \(A^{I}\) to reduce
the item \(\langle u(A^{I}u)v\rangle_{e}\). Notice that, this
procedure will not affect the critical state. Based on the
definition of \(\langle u(A^{I}u)v\rangle_{e}\), we define an
inhibitory factor \(\delta_i\) for each excitatory node \(i\) as
follows:
\begin{equation}
\delta_i=v_{i}u_{i}\sum_{k}\sum_{j}A_{ij}^{IE}A_{jk}^{EI}u_{k}.\label{E19}
\end{equation}

We now explain the meaning of the inhibitory factor \(\delta_i\). In
the case of weak stimulus, the influence of inhibitory nodes is
relatively passive. According to the rule, only the inhibitory nodes
are activated can they release inhibitory signals. Thus in this
condition, excitatory nodes actually have two effects: activating
their excitatory neighbors to transmit excitations and inhibitory
neighbors to release inhibitory signals. The later effect is
implemented via the inhibitory paths, which start from excitatory
nodes and end in E-layer through one step in I-layer, see
Fig\(\ref{illustration}(b)\). For the inhibitory paths which start
at node \(i\) and ends at node \(k\), the interaction strength is
\(\sum_{j}A_{ij}^{IE}A_{jk}^{EI}\). Because in stationary state,
excitatory nodes have different active probabilities, each
inhibitory path should have its own weight. Here we set the weight
as \(u_iu_k\), reflecting the active probability of node \(i\) and
\(k\). After summarizing the weighted interaction strength of
inhibitory paths starting from node \(i\), we have
\(u_i\sum_{k}\sum_{j}A_{ij}^{IE}A_{jk}^{EI}u_k\). Lastly, we
multiply it with \(v_i\) to reflect the active probability of node
\(i\), obtaining the definition of inhibitory factor \(\delta_i\).
Therefore, \(\delta_i\) can be interpreted as the ability of node
\(i\) to cause inhibition in stationary state. In the procedure of
enhancing dynamic range, we only need to cut out the nodes with
large \(\delta_i\) from I-layer.

\begin{figure}
  \includegraphics[width=3.5in]{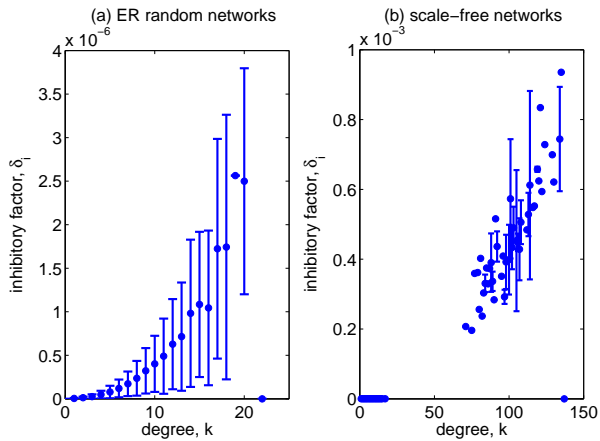}\\
  \caption{Inhibitory factor \(\delta_i\) versus node degree \(k\).
(a), simulation on ER random networks with node number \(N=10^4\),
mean degree \(\langle k\rangle=10\), and \(f_e=0.8\). (b),
simulation on BA scale-free networks with node number \(N=10^4\),
mean degree \(\langle k\rangle=4\), and \(f_e=0.8\). The solid dots
represent the mean values and the vertical lines show the standard
deviations.}\label{delta_deg}
\end{figure}

Now we check the relationship between the inhibitory factor and
degree. In Fig.\(\ref{delta_deg}\) we present the simulation results
of both ER random networks and BA scale-free networks. The solid
dots represent the mean value and the vertical lines show the
standard deviation. For ER random networks, nodes with small degree
always have small \(\delta_i\). However, for nodes with large
degree, there are huge fluctuations in \(\delta_i\). We quantify
these fluctuations with standard deviations. As can be seen, the
standard deviations are comparable with mean values, making the
ranges of inhibitory factor for different degrees overlap with each
other. More simulations with different network sizes show that these
huge relative fluctuations also exist for other system sizes. As for
scale-free networks, the values of \(\delta_i\) are divided into two
groups: for most nodes with small degree, their inhibitory factors
are negligible; whereas, highly-connected nodes have larger
\(\delta_i\), which also shows great fluctuations. Also, these
fluctuations are unrelated to system size. It can be seen there is
no clear relations between inhibitory factor and degree.
Consequently, it is unreasonable to just cut out the nodes with high
degree from the I-layer.

We give an analysis of improvement of dynamic range. By
Eq.\((\ref{E18})\) and definition of dynamic range, we have the
upper bound on the improvement in dynamic range
\begin{equation}
\Delta_u=10\log(1+\frac{\langle
u(A^Iu)v\rangle_e}{(n-\frac{1}{2})\langle
u^2v\rangle_e}).\label{E20}
\end{equation}

For ER random networks, we can simplify the calculation as follows.
For E-I networks with mean degree \(K\), on average, each excitatory
node has \(f_iK\) edges pointing to I-layer, and each inhibitory
node has \(f_eK\) edges pointing back to E-layer. So we approximate
\(\sum_jA^{IE}_{ij}A^{EI}_{jk}\approx f_iKS\cdot f_eKS/N_e\).
Substituting this term in Eq.\((\ref{E19})\), we have
\begin{equation}
\delta_i\approx v_iu_if_ef_iK^2S^2\langle u\rangle_e.\label{E21}
\end{equation}
Then we use Eq.\((\ref{E20})\) to obtain
\begin{equation}
\Delta_u=10\log(1+\frac{f_ef_iK^2S^2\langle
u\rangle_e}{N_e(n-1/2)\langle u^2v\rangle_e}).\label{E22}
\end{equation}
As for scale-free networks, we can check the upper bound through
numerical calculations.

\begin{figure}
  \includegraphics[width=3.5in]{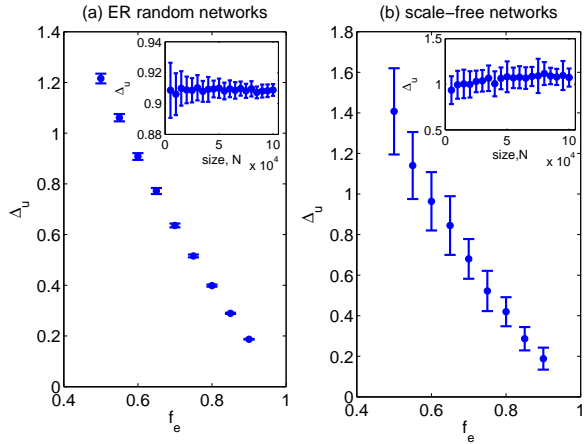}\\
  \caption{Upper bound \(\Delta_u\) versus \(f_e\). Networks are at critical state. The solid dots are mean values and vertical lines show standard deviations. (a), simulation on ER random networks with node number \(N=10^4\),
mean degree \(\langle k\rangle=20\), and \(f_e=0.6\). (b),
simulation on BA scale-free networks with node number \(N=10^4\),
mean degree \(\langle k\rangle=4\), and \(f_e=0.6\). Insets show the
relationship between \(\Delta_u\) and system size \(N\) for each
case.}\label{upper bound}
\end{figure}

Fig.\(\ref{upper bound}\) shows the simulations of upper bound
\(\Delta_u\). For both cases, \(\Delta_u\) decreases with the
increasing fraction of excitatory nodes \(f_e\). The only difference
is that scale-free networks present larger fluctuation, which cannot
affect the decreasing trend. In the insets, we explore the effect of
system size on upper bound. It can be seen for both cases,
\(\Delta_u\) is not influenced by system size. Also, the scale-free
networks show larger fluctuation. Another conclusion we get from
Fig.\(\ref{upper bound}\) is that the improvement is small compared
to the original dynamic range. This can be explained as follows.
Firstly, \(A^{IE}\) and \(A^{EI}\) are sparse, so there are not much
inhibitory paths exist. This makes the term
\(\sum_jA^{IE}_{ij}A^{EI}_{jk}\) very small. Therefore, by
Eq.\((\ref{E19})\) and \((\ref{E20})\), \(1+\frac{\langle
u(A^Iu)v\rangle_e}{(n-\frac{1}{2})\langle u^2v\rangle_e}\) is close
to \(1\), which makes \(\Delta_u\) small. Secondly, in the critical
state, each excitatory node can only excite a small number of
inhibitory nodes. Because the function of inhibitory nodes is
passive (i.e. only they are excited can they exert inhibitory
impact), this weakens the influence of inhibitory nodes.
Consequently, cutting out the inhibitory nodes can only enhance
dynamic range a little. In fact, by this method, we cannot change
the functional form \(\eta\propto F^2\). We can only change the
constant parameter.

\begin{figure}
  \includegraphics[width=3.5in]{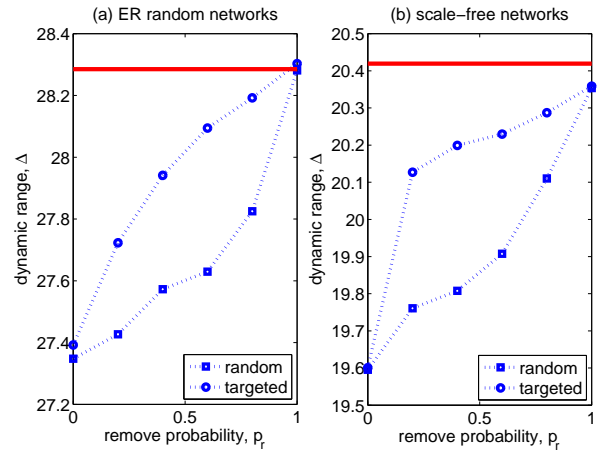}\\
  \caption{Dynamic range \(\Delta\) versus remove probability \(p_r\) for random remove and
targeted remove. The networks are at critical state. Data in two
cases are displayed with different symbols. (a), simulation on ER
random networks with node number \(N=10^4\), mean degree \(\langle
k\rangle=20\), and \(f_e=0.6\). (b), simulation on BA scale-free
networks with node number \(N=10^4\), mean degree \(\langle
k\rangle=4\), and \(f_e=0.6\). The red line is the upper bound of
improvement.}\label{remove}
\end{figure}

To check the effect of our method, we perform a comparison of random
remove and targeted remove. Denote \(p_r\) as the fraction of nodes
which are cut out from I-layer. In random remove, we choose the
nodes randomly, while in targeted remove we select the nodes with
top \(p_r\) percent inhibitory factors. In Fig.\(\ref{remove}\), we
can see for both cases, dynamic range of targeted remove increases
faster than that of random remove. However, the improvement is
relatively small compared with the original dynamic range.

\section{\label{sec6}Conclusions and Discussions}

In this paper, we propose an excitatory-inhibitory excitable network
model. To analysis the criticality and dynamic range of this model,
we divide this network into two layers: E-layer which consist of
excitatory nodes and I-Layer which only contains inhibitory ones.
Based on it, we give a theoretical analysis on random networks and
scale-free networks respectively. It is proved that, the critical
state occurs at \(\sigma_c=1/f_e\) for random networks and the
dynamic range is maximized at criticality. As for scale-free
networks, the phase transition happens when the largest eigenvalue
of the E-layer's weighted adjacency matrix \(A^{EE}\) is just one.
Similarly, the dynamic range is also optimized at critical point. It
is interesting that the critical point is not affected by the
I-layer and the links between these two layers. Then, we discuss the
method to enhance dynamic range. Based on the analysis of mutual
effects, we propose an inhibitory factor \(\delta_i\) for each
excitatory node to quantify their inhibitory ability. By cutting out
the excitatory nodes with high inhibitory ability from I-Layer, the
dynamic range can be further improved. However, the improvement is
relatively small. We give an analysis to the upper bound on
improvement in dynamic range and explain why the enhancement is
small.

For further study, it is meaningful to investigate networks with
heterogenous connectivity strength \cite{GHI}. Meanwhile, the study
of real-world neural or sensor networks is also of great
significance for the applications of the theory.

\begin{acknowledgments}
This work is partially supported by the International Cooperation
Project of The Ministry of Science and Technology of the People's
Republic of China (No. 2010DFR00700).
\end{acknowledgments}

\nocite{*}

\bibliography{apssamp}% Produces the bibliography via BibTeX.

\begin{thebibliography}{}
\bibitem{WS}
D.J. Watts and S.H. Strogatz, Nature (London) \textbf{393}, 440
(1998).
\bibitem{AB}
R. Albert and A.L. Barab\'{a}si, Science \textbf{286}, 509 (1999).
\bibitem{KC}
O. Kinouchi and M. Copelli, Nat. Phys. \textbf{2}, 348 (2008).
\bibitem{BS}
E. Bullmore and O. Sporns, Nature (London) \textbf{10}, 186 (2009).
\bibitem{PV}
R. Pastor-Satorras and A. Vespignani, Phys. Rev. Lett. \textbf{86},
3200 (2001).
\bibitem{CFL}
C. Castellano, S. Fortunato, and V. Loreto, Rev. Mod. Phys.
\textbf{81}, 2, (2009).
\bibitem{YR}
V.M. Yakovenko and J.B. Rosser Jr., Rev. Mod. Phys. \textbf{81},
1703 (2009).
\bibitem{PTZ}
S. Pei, S. Tang, X. Zhang, Z. Liu, and Z. Zheng, Physica A
\textbf{391}, 2023, (2012).
\bibitem{BP}
J.M. Beggs and D. Plenz, J. Neurosci \textbf{23}, 11167, (2003).
\bibitem{BP1}
J.M. Beggs and D. Plenz, J. Neurosci \textbf{22}, 5216, (2004).
\bibitem{SMTM}
J. Soriano, M.R. Martinez, T. Tlusty, and E. Moses, Proc. Natl.
Acad. Sci. USA \textbf{105}, 13758, (2008).
\bibitem{BSMT}
I. Breskin, J. Soriano, E. Moses, and T. Tlusty, Phys. Rev. Lett.
\textbf{97}, 188102 (2006).
\bibitem{ZK}
D.H. Zanette and M. Kuperman, Physica A \textbf{309}, 445 (2002).
\bibitem{FTV}
L.M. Fern\'{a}ndez-Carrasco, H. Terashima-Mar\'{i}n, and M.
Valenzuela-Rend\'{o}n, IEEE International Conference on Systems, Man
and Cybernetics, 1181 (2008).
\bibitem{CROK}
M. Copelli, A.C. Roque, R.F. Oliveira, and O. Kinouchi, Phys. Rev. E
\textbf{65}, 060901 (2002).
\bibitem{CORK}
M. Copelli, R.F. Oliveira, A.C. Roque, and O. Kinouchi,
Neurocomputing, \textbf{65}, 691 (2005).
\bibitem{CC}
M. Copelli and P.R.A. Campos, Eur. Phys. J. B \textbf{56}, 273
(2007).
\bibitem{WXW}
A.C. Wu, X.J. Xu, and Y.H. Wang, Phys. Rev. E \textbf{75}, 032901
(2007).
\bibitem{LSR}
D.B. Larremore, W.L. Shew, and J.G. Restrepo, Phys. Rev. Lett.
\textbf{106}, 058101 (2011).
\bibitem{LSOR}
D.B. Larremore, W.L. Shew, E. Ott, and J.G. Restrepo, Chaos
\textbf{21}, 025117 (2011).
\bibitem{AST}
Y. Adini, D. Sagi, and M. Tsodyks, Proc. Natl. Acad. Sci. U.S.A.
\textbf{94}, 10426 (1997).
\bibitem{PT}
C. Park and D. Terman, Chaos \textbf{20}, 023122 (2010).
\bibitem{VRA}
T.P. Vogels, K. Rajan, and L.F. Abbott, Annu. Rev. Neurosci.
\textbf{28}, 357 (2005).
\bibitem{FE}
S.E. Folias and G.B. Ermentrout, Phys. Rev. Lett. \textbf{107},
228103 (2011).
\bibitem{LSSA}
V. Litvak, H. Sompolinsky, I. Segev, and M. Abeles, J. Neurosci.
\textbf{7}, 3006 (2003).
\bibitem{SN0}
M.N. Shadlen and W.T. Newsom, Curr. Opin. Neurobiol. \textbf{4}, 569
(1994).
\bibitem{KAK}
J. Kremkow, A. Aertsen, and A Kumar, J. Neurosci. \textbf{47}, 15760
(2010).
\bibitem{VA}
T.P. Vogels and L.F. Abbott, Nat. Neurosci. \textbf{12}, 483 (2009).
\bibitem{SN}
M.N. Shadlen and W.T. Newsome, J. Neurosci. \textbf{10}, 3870
(1999).
\bibitem{B}
N. Brunel, J. Comput. Neurosci. \textbf{8}, 183 (2000).
\bibitem{OL}
M. Okun and I. Lampl, Nat. Neurosci. \textbf{11}, 535 (2008).
\bibitem{WC}
W.B. Wilent and D. Contreras, Nat. Neurosci. \textbf{8}, 1364
(2005).
\bibitem{KT}
M. Kurant and P. Thiran, Phys. Rev. Lett. \textbf{96}, 138701
(2006).
\bibitem{M}
C.R. MacCluer, SIAM Rev. \textbf{42}, 487 (2000).
\bibitem{GHI}
C.V. Giuraniuc, J.P.L. Hatchett, J.O. Indekeu, M. Leone, I.
P\'{e}rez Castillo, B. Van Schaeybroeck, and C. Vanderzande, Phys.
Rev. Lett. \textbf{95}, 098701 (2005).



\end{thebibliography}

\end{document}